\documentclass{elsart}

 \usepackage{epsfig}

\usepackage{amssymb}
\begin{document}
\begin{frontmatter}


\title{ A Monte Carlo Investigation of the Hamiltonian Mean Field Model  }


\author
{Alessandro Pluchino}, \noindent \ead{alessandro.pluchino@ct.infn.it}
\author
{Giuseppe Andronico},  \ead{giuseppe.andronico@ct.infn.it}
\author
{Andrea Rapisarda} \ead{andrea.rapisarda@ct.infn.it}

\address {Dipartimento di Fisica e Astronomia,  Universit\'a di Catania,\\
CACTUS Group and INFN sezione di Catania,  Via S. Sofia 64, I-95123 Catania, Italy}

\begin{abstract}


We present a Monte Carlo numerical investigation of the Hamiltonian Mean Field (HMF) model. We  begin by discussing  canonical Metropolis Monte Carlo calculations, in order to check the  caloric curve of the HMF model and study finite size effects. In the second part of the paper, we  present
numerical simulations obtained by means of a modified  Monte Carlo procedure   with the aim  to test the stability of those states at minimum temperature and zero magnetization (homogeneous Quasi Stationary States) which exist  in the condensed phase of the model just below  the critical point. For energy densities smaller than  the limiting value $U\sim 0.68$, we find 
 that these states are unstable confirming   a recent  result on the Vlasov stability analysis applied to the HMF model. 

\end{abstract}

\begin{keyword}
Hamiltonian spin model; Phase transitions; Montecarlo method; 
\PACS  75.10.Hk, 05.70.Fh, 02.70.Uu
\end{keyword}
\end{frontmatter}

\section{Introduction}
\label{intro}

The Hamiltonian Mean Field (HMF) model, introduced in ref. \cite{ruffo,leshouches2} is a classical many-body model of fully-coupled rotators, that has recently raised a great interest for its connections to nonextensive thermodynamics \cite{pre,cmt,leshouches1,tsagel,cho,plud,celia1} and glassy systems \cite{plu1,plu2}.
It can be seen as a system of  $N$  planar classical inertial spins  
${\stackrel{\vector(1,0){8}}{s_i}}=(cos\theta_i, sin\theta_i)$ 
which interact  through an infinite-range potential  or equivalently as $N$ interacting particles moving on the unit circle 
\cite{leshouches2}. 
Indicating  with $K$ the kinetic energy and with $V$ the potential energy, the Hamiltonian
can be written as
\begin{equation}
\label{eq1}
        H= K+ V
= \sum_{i=1}^N  {{p_i}^2 \over 2} +
  {1\over{2N}} \sum_{i,j=1}^N  [1-cos(\theta_i -\theta_j)]~~,
\end{equation}
\noindent
where $-{\pi}\le {\theta_i}<{\pi}$ is the $ith$ angle and $p_i$ the 
corresponding  conjugate momentum. 
The degree of clustering of the system can be expressed through the usual order parameter M, defined as  
\begin{equation}
\label{eq2}
M = {1\over{N}} | \sum_{i=1}^N 
\stackrel{\vector(1,0){8}}{s_i} |~~~~.
\end{equation}
The model can be solved exactly in the canonical ensemble formalism and exhibits a second-order phase transition from a low-energy 
condensed (ferromagnetic) phase with magnetization $M\ne0$,  
to a high-energy  homogeneous one (paramagnetic), with $M=0$.
The dependence of the temperature $T={2<K>\over N}$ on the energy density $U = E/N$ at equilibrium
is given by the following  {\em caloric curve} \cite{ruffo,leshouches2}
\begin{equation}
U = {T \over 2} + {1\over 2} \left( 1 - M^2 \right) ~.
\end{equation}
The
critical point is at energy density $U_c=0.75$ and the 
corresponding  critical temperature is $T_c=0.5$. 
The dynamics of HMF can be investigated by numerical integration of
 the equations of motion at constant energy. 
Starting the 
system with out-of-equilibrium initial conditions, for example adopting 
the so-called M1
initial conditions (i.e. considering $\theta_i=0$  for 
all $i$ - so that $M(t=0)=1$ - and velocities uniformly distributed), 
the results of the microcanonical molecular dynamics (MD) simulations show a disagreement with the canonical prediction in a region of energy values $0.5<U<U_c$ \cite{pre,leshouches1,plud}. Here,
for a transient regime whose length depends on the  size N,
the system remains trapped in anomalous metastable quasi-stationary states (QSS)
at a temperature lower then the canonical equilibrium one. After such a transient, for a finite size N, the system slowly relaxes towards Boltzmann-Gibbs (BG) equilibrium,
showing aging and power-law correlations \cite{plud}. Moreover, in the thermodynamic limit, if the infinite time limit  is considered after $N\rightarrow \infty$, the quasi-stationary states become stable and the QSS regime can be considered as a true non-canonical equilibrium phase of the model
\cite{pre}.
In the last years many investigations have been  performed  in order to explain the nature of the dynamically-generated anomalies of the QSS regime. The most promising scenarios seem to be  Tsallis nonextensive statistical mechanics framework \cite{pre,leshouches1,plud} and the glassy-like weak ergodicity breaking description \cite{plu1,plu2}. Both of these scenarios indicate  that, during the QSS regime, the system remains trapped in a very complex (fractal) region of the phase space, which hinders the complete  visit of the  whole  a-priori accessible phase space.
\\
In this paper we present a Monte Carlo  study of the HMF model. In fact an extensive Monte Carlo investigation is missing in the literature and only in refs. \cite{gross,salaz} some calculations for very small sizes were discussed. The present paper  is divided into two sections. In the first one, by means of a standard  Metropolis algorithm, we reproduce the canonical equilibrium caloric curve of the model and study finite size effects close to the critical point where fluctuations are larger
and simulations more delicate. In the second section, we modify the standard algorithm in order to perform a sampling over the constant energy hypersurface of phase space, looking for those states with  minimum temperature (kinetic energy),  not necessarily
Boltzmann-Gibbs equilibrium states. Below  the critical energy $U_c=0.75$ density, these spatially homogeneous $(M=0)$ states coincides with the microcanonical non-equilibrium QSS for energy greater than  $U\sim0.68$. With our modified Monte Carlo algorithm we found that below this limiting value, as also showed in a recent paper \cite{celia2} by means of a nonlinear  stability test, the homogeneous states are effectively unstable.
We discuss  these results comparing them with  the out-of-equilibrium caloric curve calculated
using molecular dynamics and with the absolute minimum temperature curve.

\section{ Metropolis Monte Carlo simulations }
\label{metropolis}

The  Metropolis Monte Carlo  algorithm
is a well-known general method \cite{binder} for computing the canonical equilibrium statistical expectation values by means of 
a weigthed random sampling of the possible microstates.
The algorithm usually generates a Markov chain of configurations,
for which the probability of having a given configuration $C_n$ depends only
on $C_{n-1}$ and not on the previous history of the system.
Given a configuration $C_n$, one extracts a new trial configuration $C'$ with
a random algorithm characterized by a simmetric transition probability, in order to satisfy the detailed balance condition and to minimize the free energy.
For example, in the Ising model the configuration $C'$
could be the configuration $C_n$ in which a random chosen spin has been flipped.
More generally, the configuration $C'$ could be obtained from $C_n$ by changing at random
the state of the $ith$ spin of the system considered.
Then, if the equilibrium distribution is $exp[-\beta H(C)]$, 
where $H(C)$ is the Hamiltonian of the system and $\beta$ the inverse of the temperature T (which remains constant), one computes $H(C')-H(C_n)\equiv \Delta H$ and uses
the extraction of a uniformly distributed
random number $r$ in the interval [0,1] in order to accept the new configuration, according to the following rules
\[
C_{n+1}=C_n~~~~~~~if~~~exp[-\beta \Delta H]<r
\]
\begin{equation}
C_{n+1}=C'~~~~~~~if~~~exp[-\beta \Delta H]>r~~,
\label{delta}
\end{equation}
By means of this technique
one can generate a sequence of configurations from
which, after an opportune thermalization transient, it is possible
to  get the canonical equilibrium properties of the system. In fact, 
if $N_{iter}$ is the number of iterations of the algorithm,
the expectation value of any observable quantity $A(C)$ can be calculated 
using the identity
\begin{equation}
\label{identity}
\lim_{N_{iter}\rightarrow \infty} \frac{1}{N_t} \sum_{n=1}^{N_{iter}} A(C_n) = \int d\mu(C)A(C) ,
\end{equation}
with
\begin{equation}
d\mu(C) \sim exp[-\beta H(C)],~~~~~~~\int d\mu(C)=1~~.
\end{equation}

%

\begin{figure}
\begin{center}
\epsfig{figure=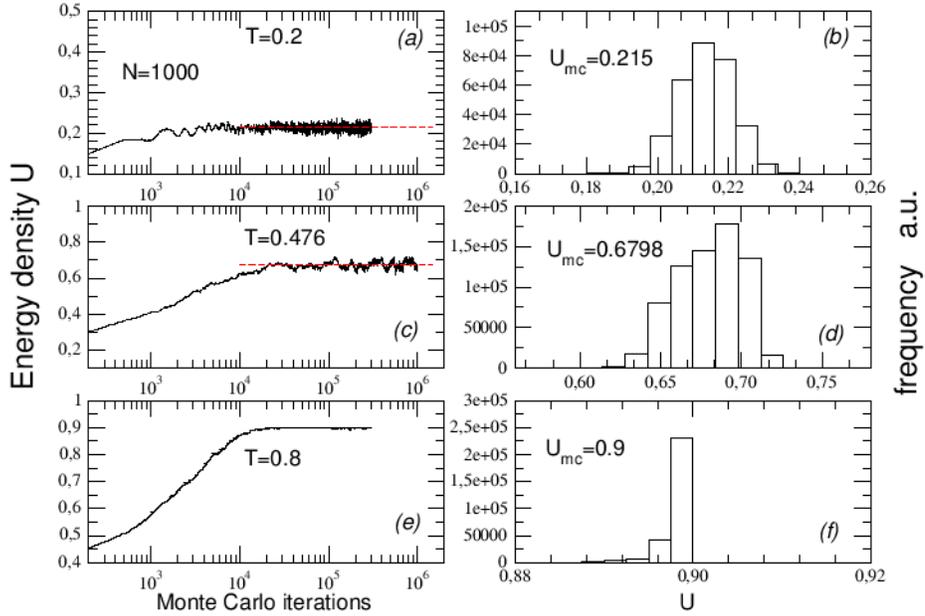,width=13truecm,angle=0}
\end{center}
\caption{For the case $N=1000$ we plot in  panels  (a), (c), (e)   the standard Metropolis Monte Carlo evolution of the energy density U and the relative histograms,  in panels (b), (d), (f) for three different values of temperature, i.e.  $T=0.2,0.476,0.8$,  which are , respectively, below, near and above the phase transition.
}
\end{figure}

At variance 
with usual statistical models, the HMF model has  also a kinetic term. Thus  
in order to use the standard Metropolis algorithm and calculate  the caloric curve of the HMF model,  we 
 fixed the parameter $\beta = 1/T$, in this way  the kinetic energy $K=NT/2$ of the system of N rotators is
 fixed as well and remains constant for all the simulation.   We  started from 
  initial conditions with  all the angles equal to zero, i.e. $M(t=0)\sim 1$.
Then ,  in the configuration $C_n$, we changed the angle $\theta_k$ of a given small quantity (choosen properly in order to satisfy the detailed balance condition), so that, in the new configuration $C'$, we have $\theta'_k = \theta_k + \Delta \theta_k$.
We computed the corresponding variation $\Delta V$ in the potential energy and since $\Delta K=0$ we have $\Delta H = \Delta V$.
Following the rule of eq. (\ref{delta}) we finally accepted or not the new configuration.
 After a termalization transient,  we started to compute the average energy density $U=H/N$ by means  of eq.(\ref{identity}).

\begin{figure}
\begin{center}
\epsfig{figure=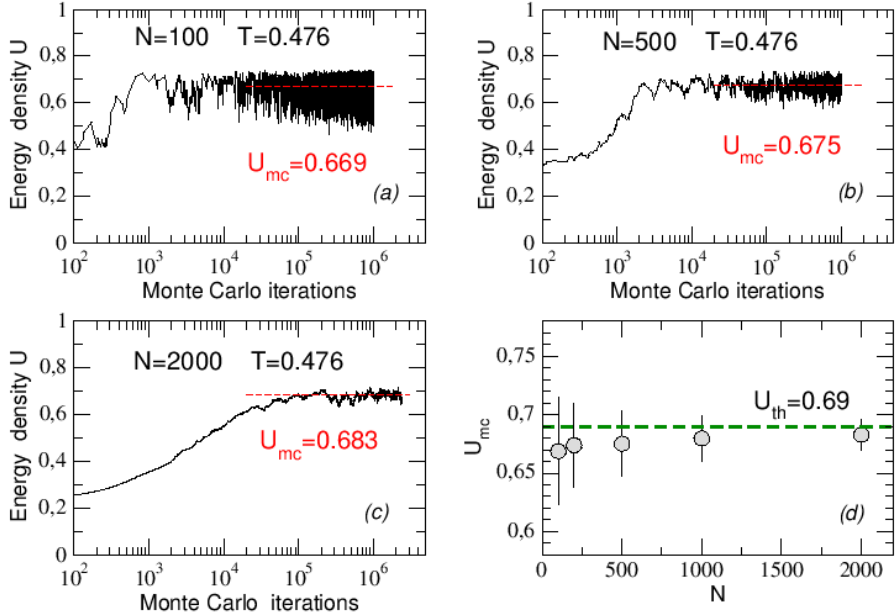,width=13truecm,angle=0}
\end{center}
\caption{For the case $T=0.476$ which corresponds to the well studied energy density value $U=0.69$, 
we plot in panels (a), (b) and (c)   the standard Metropolis Monte Carlo evolution  for different $N$  in order 
to investigate finite size effects. The average value $U_{mc} $ is also reported and shown as dashed line.
In panel (d) we plot the values of the energy density obtained as a function of the system size. A clear convergence towards the theoretical prediction $U_{th}$ is evident together with a decrease of the error bar which is given by the corresponding standard deviation extracted from the numerical fluctuations around the average value.  
}
\end{figure}


\begin{figure}
\begin{center}
\epsfig{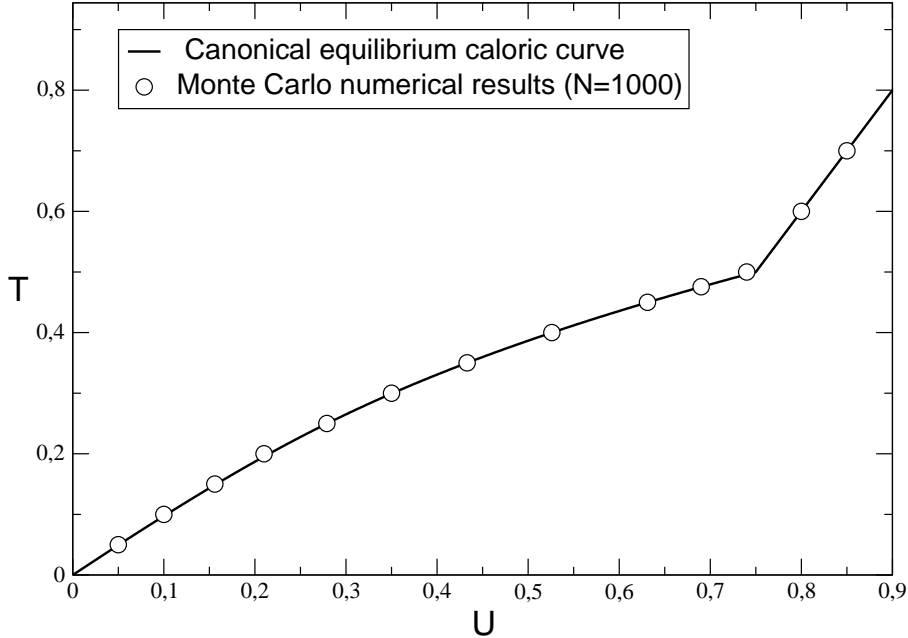}
\end{center}
\caption{The  numerical results of the standard Metropolis simulations performed for a system of N=1000 rotators (circles) are plotted in comparison with the analytical  canonical caloric curve, corresponding to the Boltzmann-Gibbs equilibrium (full curve).
}
\end{figure}

Let us discuss in the following the numerical results obtained with this standard Monte Carlo algorithm.
In Fig.1,panels (a), (c) and (e),  we plot    Metropolis simulations performed for a system of N=1000 rotators and  three different values of temperatures. We considered in particular the values 
  $T=0.2, 0.476, 0.8$ which are, respectively, below, near and above the phase transition crtitical temperature $T_c=0.5$.
   After a transient time, the simulations converge to a plateau and fluctuate  around an average value. 
We report in panels (b), (d) and (f)   the relative histograms with the average Monte Carlo values 
$U_{mc}$ obtained. As expected,  fluctuations are larger for temperatures close to the critical one. 

In order to study finite size effects and fluctuations we discuss the evolution for different system sizes 
at a temperature  just below $T_c$.  We considered the case $T=0.476$ which should correspond to the 
well known energy density $U=0.69$. In Fig.2 (a), (b) and (c), we show the Monte Carlo evolution for this case  for different sizes of the system. While  fluctuations are quite consistent for $N=100$, they diminish as expected with the size of the system. Also the value to which the simulation finally converges gets closer to the theoretical prediction the bigger the size of the system. This is evident in panel (d) of the same 
figure where the values obtained for different numbers of rotators together with the relative standard deviations 
are plotted in comparison with the theoretical expected one.  

Finally, in Fig.3,  we plot the average energy densities obtained with the Monte Carlo simulations  for a wide range of temperature values. The Monte Carlo points are compared  with the analytical canonical  curve (full curve)
corresponding to the Boltzmann-Gibbs equilibrium \cite{ruffo,leshouches2}. The agreement is very 
good for this energy size (N=1000).  As expected at equilibrium, where   ergodicity should  be  satisfied 
no anamalies are found.  A similar results was obtained for microcanonical Monte Carlo simulations 
\cite{gross,salaz}. 
On the other hand, as discussed
in refs. \cite{pre,cmt,leshouches1,tsagel,plud},  the origin of the anomalies is in the dynamics which
induces frustration and nonergodicity for a transient time - before complete equilibration -  which diverges with $N$.  In order to try
to shed more light on these metastable states we modified the standard Monte Carlo algorithm as explained
in the next section. 
\\

\section{Monte Carlo optimization at constant Energy}
\label{newmc}

\subsection{The algorithm}
We mentioned in the introduction that the so-called QSS, i.e. the out-of-equilibrium metastable states that emerge from the microcanonical molecular dynamics  simulations, become stationary in the thermodynamic limit and lie at temperature lower than the canonical equilibrium one. More precisely, up to $U\sim 0.68$, the QSS points lie on the extension, in the condensed phase, of the high temperature line of the caloric curve (this line is the geometric place of states with M=0, maximum potential energy and minimum kinetic one). This means that the homogeneous quasi-stationary non-equilibrium states with zero magnetization should be stable only up to those limiting values, but not below.
In a recent paper on the HMF model \cite{celia2}, the authors apply a nonlinear stability criterion (a modification of that one originally proposed in ref. \cite{yama}) to a selected set of spatially homogeneous solutions of the Vlasov equation, which describes the continuum limit of the Hamiltonian Mean Field model. 
Actually these solutions are qualitatively very similar to the zero magnetization QSS arising from the microcanonical simulations with M1 initial conditions and also the results of the stability test found in \cite{celia2} were consistent with the numerical evidence of the disappearance of the homogeneous QSS family below $U\sim 0.68$.
\\
With the aim  to study these anamalous states without using molecular dynamics tecniques, but only performing a sampling over the constant energy density hypersurface in  phase space, we modified the standard MC Metropolis algorithm adopted in section \ref{metropolis}.
Our purpose was not that one of  finding Boltzmann-Gibbs  equilibrium configurations,   on the other hand we 
wanted to  select   those out-of-equilibrium configurations which minimize the temperature (i.e. the kinetic energy) of the system at constant total energy, in order to look for  homogeneous QSS's close to the critical point  and study their eventual stability. 
The problem is an optimization-like one, i.e. using a Monte Carlo procedure, one  wants to find the states at minimal temperature with the constraint of total energy conservation.

The new algorithm  we have adopted obeys the following rules:

\begin{enumerate}

\item  It starts from  initial conditions with fixed magnetization  and  momenta uniformly distributed. Then randomly changes the momentum $p_k$ in the configuration $C_n$ so that $p_k + \Delta p_k = \Pi$; thus, in the new configuration $C'$, one has
\[
p'_i=p_i + (\Pi - p_i)\delta_{i,k}~~~~~~~with ~~~i=1,....,N;
\]

\item  it computes the corresponding  variation in the kinetic energy
\[
K'= \frac{1}{2} \sum_{i=1}^N [p_i^2 + (\Pi^2 - p_i^2)\delta_{i,k}]= K+\frac{1}{2} \Pi^2 - \frac{1}{2} p_k^2~ \Longrightarrow 
\]
\[
\Longrightarrow ~\Delta K= \frac{1}{2} \Pi^2 - \frac{1}{2} p_k^2
\]

\item  then the $n+1$  configuration is
\[
C_{n+1}=C_n~~~~~~~if~~~exp[-2\Delta K]<r
\]
\begin{equation}
C_{n+1}=C'~~~~~~~if~~~exp[-2\Delta K]>r
\end{equation}
where $r$ is the usual random number chosen uniformly in the interval [0,1];

\item  finally, if the new configuration $C'$ is accepted, it calculates the new value $\Theta$ of the angle $\theta_k$ by
means of the variation of the potential energy $\Delta V=-\Delta K$, knowing that
\begin{equation}
V'= V + \Delta V =\frac{1}{2N} \sum_{i,j=1}^N [1-\cos(\theta'_i - \theta'_j)]
\label{V'}
\end{equation}
and
\[
\theta'_i=\theta_i + (\Theta - \theta_i)\delta_{i,k}~~~~~~~~~~i=1,....,N~.
\]
After some algebra, one finally obtains the following equation for $\Theta$
\begin{equation}
a~cos\Theta + b~sin\Theta = c
\label{eq_theta}
\end{equation}
where
\[
a= \sum_{i \neq k}^N cos\theta_i  ~~~~~b= \sum_{i \neq k}^N sin\theta_i  ~~~~~c= \sum_{i \neq k}^N cos(\theta_i - \theta_k) + N\Delta K~~.
\]
%

\begin{figure}
\begin{center}
\epsfig{figure=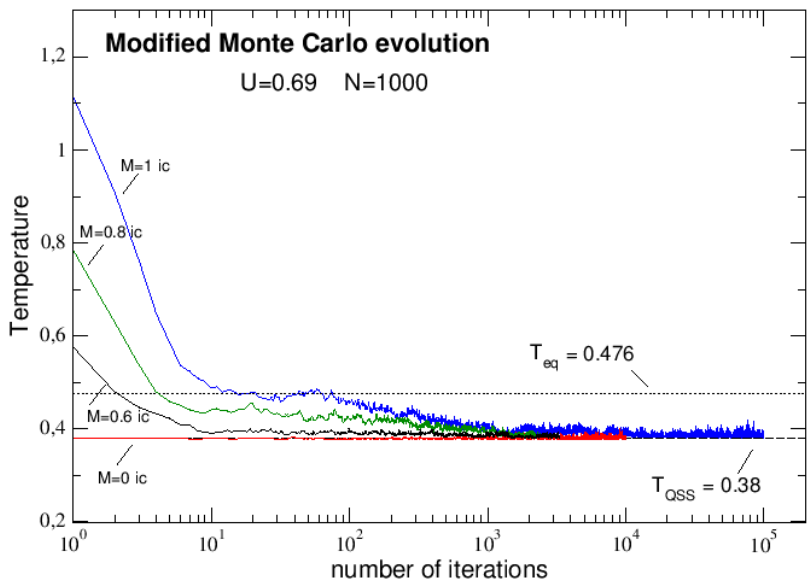,width=12truecm,angle=0}
\end{center}
\caption{We plot four examples of temperature evolution for the energy density $U=0.69$ 
and $N=1000$ obtained with the modified Monte Carlo procedure at constant energy. 
The  curves refer to various initial conditions with uniform distribution of momenta and 
magnetization M=1,0.8,0.6,0, see text.
We report also the values of temperature for the canonical equilibrium $T_{eq}=0.476$ and for the 
QSS regime in the infinite size limit $T_{QSS}=0.38$.
}
\end{figure}

\noindent
The solutions of eq.(\ref{eq_theta}) are
\begin{equation}
\Theta = \pm arccos~R^{\pm}
\label{theta}
\end{equation}
with
\begin{equation}
R^{\pm}= \frac{ac \pm bD}{a^2 + b^2}, ~~~ -1 \le R^{\pm} \le 1 , ~~~D^2=a^2 + b^2 - c^2 \ge 0~.
\label{constr}
\end{equation}
Thus, from eq.(\ref{theta}), one can calculate the new value of the $k$th angle, taking into account the new constraints (\ref{constr}). Then one can repeat the algorithm for the new configuration $C'$ and so on, until the system reaches a stationary state. At this point, the desired expectation values (in this case the temperature of the system at fixed energy density) can be calculated by means of eq.(\ref{identity}).

\end{enumerate}

\subsection{Numerical results}

With this new Monte Carlo algorithm,  we calculated a new out-of equilibrium caloric curve for the HMF model, which corresponds to those states with  minimal temperature at constant total energy density.
\\
In our simulations we consider N=1000 rotators and  different initial conditions
with uniform distribution of momenta and different initial magnetization varying from $M= 1$  to $M=0$ \cite{plud}\footnote{The inital condition M=0 can be adopted only for
$U \ge0.5 $}.  Then we follow the system evolution until a stationary value of temperature has been reached for each energy density value considered. In Fig.4 we show for example the temperature evolution in  phase space for $U=0.69$.  The plot shows that, using this new optimization algorithm, the Boltzmann-Gibbs equibrium temperature $T_{eq}=0.476$ is not a
stable minimum, the curve for $M=1 $ stays there for a while and then escapes to reach a deeper  minimum.
Instead, as expected for this energy density, the most stable solution is always the temperature corresponding to the homogeneous QSS $T_{QSS}=0.38$, also reported in the figure.


\begin{figure}
\begin{center}
\epsfig{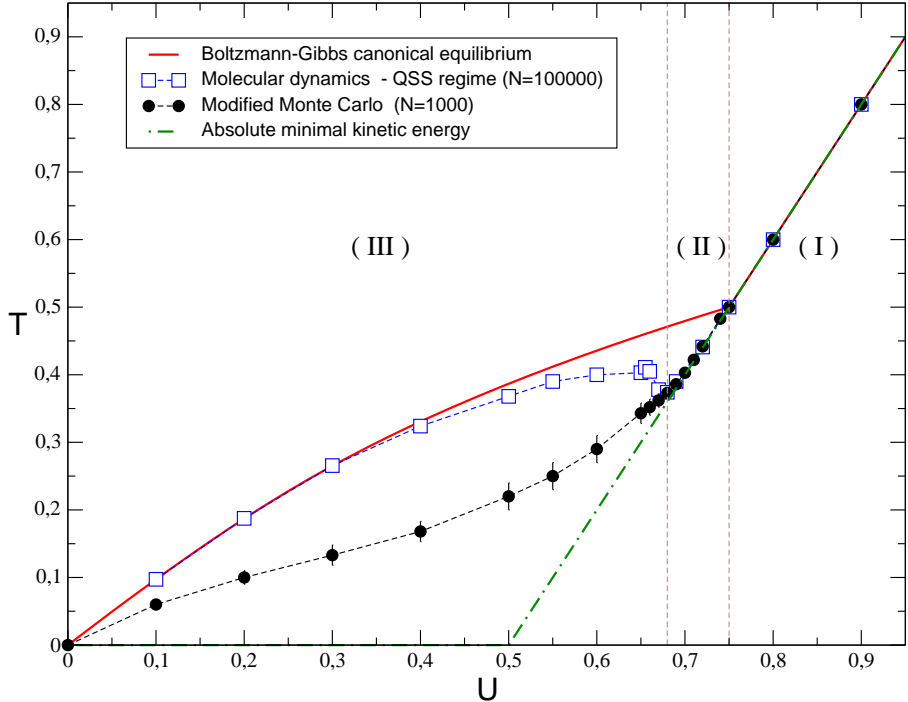}
\end{center}
\caption{The caloric curve obtained with the Monte Carlo optimization procedure (full points) is compared with the Boltzmann-Gibbs canonical prediction at equilibrium (full curve) and with the metastable QSS caloric curve (open squares) obtained by means of  the molecular dynamics  microcanonical simulations  in the QSS regime.
We report for comparison
 also the curve corresponding to the a-priori absolute minimal values of kinetic energy (dot-dashed).
We can distinguish three different regions: (I) In the homogeneous phase all the curves coincide. (II) In the energy interval
$0.68\le U \le U_c$  ($U_c=0.75 $) - marked by the two vertical dashed lines -
 the Monte Carlo curve differs from the Boltzman-Gibbs equilibrium prediction, and lies on the extension of the high temperature branch as the  microcanonical dynamical simulations. (III) Below U=0.68, while the molecular dynamics curve  rejoins again  the canonical one, the Monte Carlo curve remains at a lower (but not at its absolute minimum) temperature value for  the whole condensed phase. }
\end{figure}

In fig.5 we plot the caloric  curve obtained with this new Monte Carlo optimization approach (full points).
 The bars reported indicate the size of the fluctuations. In comparison we show also  the Boltzmann-Gibbs  canonical equilibrium prediction (full line) and   the metastable caloric curve found by means of molecular dynamics microcanonical simulations, performed for N=100000 with  M1 initial conditions, in the QSS regime (open squares) \cite{plud}.
It is helpful to divide the graph in three regions.
\\
In  region (I), i.e. in the homogeneous high temperature phase, the three curves coincide. This  means that in this phase the Monte Carlo optimization method is able to recover the expected canonical equilibrium curve. This is in agreement with the fact that here  the total energy is essentially kinetic: the potential energy is constant and equal to its  maximum value V/N=0.5.
\\
On the contrary, in region (II) - i.e. between $U_c$ and $U\sim 0.68$ - the Monte Carlo curve deviates from the canonical prediction, but coincides with the results of the molecular dynamics   simulations in the QSS regime. As in the previous region, here also the Monte Carlo temperature takes its minimum allowed value, consistently with the total energy constraint.
This fact is interesting in order to understand the nature of the QSS regime, which here
corresponds to a stable attractor of the system when  the magnetization (and the force
acting on the single spin)  is zero. We have checked that this result does not change with  the number of spins considered.
\\
Below the value  $U=0.68$ - just lower than the value  ($U=0.69$)  where the QSS dynamical anomalies and the disagreement with the canonical  prediction are more evident - the Monte Carlo curve starts to disagree also with the molecular dynamics simulations. In fact, in the region (III), while the QSS anomalies tend to disappear and the molecular dynamics curve slowly rejoins (below $U=0.5$) the canonical Boltzmann-Gibbs equilibrium prediction, the Monte Carlo curve stays below the other two curves. However the Monte Carlo result does not correspond to the absolute minimum value of temperature. In Fig.5 we report, for comparison, the  curve  (dot-dashed) corresponding to the a-priori absolute minimum value of kinetic energy. 
 The Monte Carlo curve differs from such a minimal kinetic energy  - which corresponds to $M=0$ -  around 
  the limiting value $U\sim 0.68$. This is a numerical confirmation  of the instability of the homogeneous quasi-stationary states below this value as found in the nonlinear stability test of ref. \cite{celia2}.
\\
Finally, it is interesting to observe that below the value $U=0.5$, where the kinetic energy - and therefore the temperature - could also be null, the Monte Carlo curve lies roughly in the middle between zero and the equilibrium temperature values. 
This result is not fully understood and could be originated by a competition between the two unstable  attractors 
 in  phase space.
 Further investigations in this respect  is  required.

\section{Conclusions}

In this paper we have presented  a Monte Carlo study of the HMF model. In the first part of the
paper by means of a standard  Metropolis  procedure, we were able to  reproduce the canonical caloric 
curve of the HMF model at  equilibrium and study finite size effects close to the  critical point where 
dynamical anomalies exist in the out-of-equilibrium regime.  
In the second part of the paper we studied out-of-equilibrium states by means of a Monte Carlo 
optimization technique. To this end, we have  modified    the standard Metropolis algorithm, in order to obtain a 
temperature minimization at constant energy density and look for  those states with minimal temperature, 
which  are not Boltzmann-Gibbs  equilibrium states. 
However in this  way we could study  those metastable homogeneous Quasi Stationary States found 
dinamically below the phase transition point. We found that these states 
  are stable for energy densities greater than $U\sim 0.68$. For energy densities smaller
  than this value, the zero magnetization quasi-stationary states  are not reached and a different 
  caloric curve is obtained. 
  These results confirm what found in  \cite{celia2} by means of a nonlinear stability
  analysis where the authors suggest  that this fact 
is   due to the loose of stability of the spatially homogeneous solutions of Vlasov equation. 
We hope that this work, supplying a non-dynamical point of view and 
adding new numerical results to the study of the HMF model, 
will be of help in sheding further light on the nature of 
the several intriguing aspects of this  model.

\vskip 0.25truecm
\noindent
We thank C. Anteneodo, F. Baldovin, V. Latora and  C. Tsallis for  stimulating discussions.
We  would like  to dedicate this paper to Constantino Tsallis for his 60th birthday wishing him 
a still very long and fruitful research activity.
\bigskip

\noindent


\begin{thebibliography}{00}

\bibitem{ruffo} M. Antoni  and S.Ruffo, Phys.  Rev. E {\bf 52} (1995) 2361.

\bibitem{leshouches2} For a recent review on this model see also: T. Dauxois, V. Latora, A. Rapisarda, S.  Ruffo, A. Torcini, in {\it Dynamics and Thermodynamics of Systems with Long-Range
Interactions} T. Dauxois, S. Ruffo, E.  Arimondo, M. Wilkens
Eds., Lecture Notes in Physics Vol. 602, Spinger (2002) p.458 and references therein.

\bibitem{pre} V. Latora , A. Rapisarda  and C. Tsallis, {\it
Phys. Rev. E} {\bf 64}  (2001) 056134.

\bibitem{cmt} A. Pluchino,  V. Latora,  A. Rapisarda, {\it Continuum Mechanics and Thermodynamics }  {\bf 16} (2004) 245.

\bibitem{leshouches1} C. Tsallis, A. Rapisarda, V. Latora, F. Baldovin in {\it Dynamics and Thermodynamics of Systems with Long-Range
Interactions} T. Dauxois, S. Ruffo, E.  Arimondo, M. Wilkens
Eds., Lecture Notes in Physics Vol. 602, Spinger (2002), p.140 and references therein.

\bibitem{tsagel} {\it Nonextensive Entropy: interdisciplinar ideas},
C. Tsallis and M. Gell-Mann Eds., Oxford University Press (2004).

\bibitem{cho}
 A. Cho, Science {\bf 297} (2002) 1268; Letters to the
Editors by S. Abe, A.K. Rajagopal; 
 A. Plastino;  and  V. Latora,   A. Rapisarda, A. Robledo,  {\it Science} {\bf 300} (2003) 249.

\bibitem{plud}  A. Pluchino,  V. Latora,  A. Rapisarda, {Physica D} {\bf 193} (2004) 315.

\bibitem{celia1} M.A.Montemurro, F.A.Tamarit and C.Anteneodo, {\it Phys. Rev. E} {\bf 67} (2003) 031106.

\bibitem{plu1}  A. Pluchino,  V. Latora,  A. Rapisarda, {\it Phys. Rev.  E} {\bf 69} (2004) 056113.

\bibitem{plu2}  A. Pluchino,  V. Latora,  A. Rapisarda, {\it Physica A} {\bf 340} (2004) 187.

\bibitem{gross} D.H.E.  Gross, {\it Microcanonical Thermodynamics: phase transitions in small systems}, Lecture Notes in physics, vol.66, World Scientific, Singapore 2001, pg.196-197.


\bibitem{salaz} R. Salazar, R. Toral, A.R. Plastino, {\it Physica A} {\bf 305} (2002) 144.


\bibitem{binder} D.P. Landau and K. Binder,    {\it A Guide to Monte Carlo Simulations in Statistical Physics} Cambridge University Press  (2000).

\bibitem{celia2} C.Anteneodo and Raul O.Vallejos,  {\it Physica A} (2004)  in press [cond-mat/0401195].

\bibitem{yama} Y. Yamaguchi, J.Barr\'e, F.Bouchet, T.Dauxois and S.Ruffo, {\it Physica A}  {\bf 337} (2004) 653.


\end{thebibliography}
\end{document}